\documentclass[thmsa,onecolumn,a4paper]{article}%
\usepackage{amssymb}
\usepackage{sw20jart}
\usepackage{amsmath}
\usepackage{graphicx}
\usepackage{amsfonts}%
\providecommand{\U}[1]{\protect\rule{.1in}{.1in}}
\begin{document}

\author{Bertrand Chauvineau}
\title{On gravitational radiation in Brans-Dicke gravity}
\date{Universit\'{e} C\^{o}te d'Azur, OCA, CNRS, Lagrange, France\\
(e-mail : Bertrand.Chauvineau@oca.eu)}
\maketitle

\begin{abstract}
Brans-Dicke gravity admits spherical solutions describing naked singularities
rather than black holes. Depending on some parameters entering such a
solution, stable circular orbits exist for all radius. One argues that,
despite the fact that the naked singularity is an infinite redshift location,
the far observed orbital motion frequency is unbounded for an adiabatically
decreasing radius. This is a salient difference with General Relativity, and
the incidence on the gravitational radiation by EMRI systems is stressed.
Since this behaviour survives the $\omega\longrightarrow\infty$ limit, the
possibility of such solutions is of utmost interest in the new gravitational
wave astronomy context, despite the current constraints on scalar-tensor gravity.

\end{abstract}

\baselineskip12truept

\bigskip


\bigskip

\noindent\textbf{I - Introduction}

The recently born gravitational wave (GW) astronomy opens unprecedented ways
to observe our Universe. The observation of black hole (BH) merging events,
like GW150914 and GW151226 [1][2], can only be achieved by the means of GW
observatories, as long as no electromagnetic counterpart is expected. On the
other hand, the templates that serve to modelise the time evolution of the
signal depend on the gravity theory that is supposed to govern the evolution
of the source, especially in the strong field case, like BH merging. As a
consequence, the GW astronomy is a very specific way for testing gravity
theories in the strong field regime.

So far, the observed GW events are well interpreted in the usual General
Relativity (GR) framework. Indeed, GR allows to interpret both GW150914 and
GW151226 as the final inspiraling phases and merging of BH pairs, including
the ringdown emission by the resulting BH. So far, solar system and binary
pulsar tests strongly constrain alternatives to GR in the weak (and
intermediate) field regime(s) [3]. The GW events analysis then strengthens our
confidence in GR, grounded in long standing tests. Nevertheless, many attempts
to quantize gravity (in a unified sheme of interactions or not) return a
scalar as a partner of the metric in the effective gravitational sector of the
theory [4][5][6]. This promotes Brans-Dicke (BD) and scalar-tensor (ST)
theories as serious alternatives to GR, despite the fact that there is no
experimental evidence of such a scalar so far. The credibility of such
theories is reinforced by the fact that BD/ST gravity includes GR in some
limit cases [3], and also by the fact that the Universe's expansion makes a
large family of ST theories to asymptotically behave like GR [7][8].

In any case, it is worth trying other astrophysical scenarios to interpret the
observed GW events, in both usual GR and alternative gravity frameworks. The
interest of such attempts is strengthened by the fact that GWs are expected to
reveal the existence of objects that could escape traditional
(electromagnetic) observational means. This justifies a systematic study of
the GW emission properties of speculative objects. In these lines, the link
between the ringdown phase and the existence of an event horizon is not fully
obvious. Some authors argue that the ringdown phase is not the signature of a
BH horizon, but rather of a "photonic sphere", that are associated to wormhole
structures as well, for instance [9][10]. Alternative GW sources, like the
collision of ultracompact boson stars [10], or the production of a gravastar
as the results of the merging of two ultracompact objects [11], have also been
considered. Some authors have pointed out the lack of understanding of merging
phenomena in alternative theories [12].

The aim of this paper is to spot a salient difference between GR and BD
gravity. Unlike GR, BD admits spherical solutions that exhibit a naked
singularity (NakS), rather than BH, structure. Depending on a specific
parameter related to the NakS structure, stable circular orbits exist for all
radius, with the specific property that the corresponding far observed orbital
frequency is unbounded. This last points drastically departs from GR, in which
(1) the orbital frequencies are bounded by the frequency at the innermost
stable circular orbit (ISCO), and (2) frequencies are frozen at the horizon.
Besides, one shows that these features exist for all $\omega$, and survive the
$\omega\longrightarrow\infty$ limit. The possibility of such a scenario is
then not ruled out by current constraints on gravity. Thence, if NakS turn out
to exist in our Universe, these features result in a specific signature that
carries information of the utmost relevance on the very nature of gravity.

\textit{Notations}

Unless otherwise stated, the spherical metrics hereafter considered will be
written with the help of spherical like coordinates%
\begin{equation}
ds^{2}=g_{00}dt^{2}+g_{11}dr^{2}+g_{22}\left(  d\theta^{2}+\sin^{2}\theta
d\phi^{2}\right)  . \label{spherical metric}%
\end{equation}
The geodesic equations of the metric (\ref{spherical metric}) read (the prime
representing $r$ derivation)%
\begin{align}
g_{00}\frac{dt}{dp}  &  =-E\label{geod eq}\\
2\frac{d}{dp}\left(  g_{11}\frac{dr}{dp}\right)   &  =\left(  \frac{dt}%
{dp}\right)  ^{2}g_{00}^{\prime}+\left(  \frac{dr}{dp}\right)  ^{2}%
g_{11}^{\prime}+\left(  \frac{d\theta}{dp}\right)  ^{2}g_{22}^{\prime
}\nonumber\\
g_{22}\frac{d\theta}{dp}  &  =C\nonumber
\end{align}
where $E$ and $C$ are integration constants, and $p$ is an affine parameter.
These equations admit the first integral%
\begin{equation}
g_{00}\left(  \frac{dt}{dp}\right)  ^{2}+g_{11}\left(  \frac{dr}{dp}\right)
^{2}+g_{22}\left(  \frac{d\theta}{dp}\right)  ^{2}=-\sigma^{2}
\label{geod first integral}%
\end{equation}
where $\sigma^{2}$ is a (positive) constant. In the massless particle (usually
refered to as the photonic) orbit case, one has $\sigma=0$. In the timelike
orbit case, corresponding to massive particles, $\sigma^{2}$ is positive and
will be normalized to unity by choosing the proper time as the affine
parameter (and renormalizing $E$ and $C$ accordingly).

In the case where isotropic coordinates are chosen, the metric functions
satisfy%
\begin{equation}
g_{22}=r^{2}g_{11}. \label{isotropic condition}%
\end{equation}

\bigskip

\noindent\textbf{II - ST \& BD gravity}

In this section, we first remind the general setting of ST gravity. We then
specify to the Brans's Class I spherical BD solution, presented in a form
adapted to our purpose. The detailed properties of circular orbits, that will
be relevant for the discussion on far observed frequencies discussed in the
next section, are then pointed out.

\bigskip

\noindent\textbf{II-1 - General Relativity and scalar-tensor equations}

The GR field equations are derived from the action [13]%
\begin{equation}
S_{GR}\left[  g,\Psi\right]  =\frac{1}{16\pi G}\int\sqrt{-g}\left(
R-2\Lambda\right)  d^{4}x+S_{mat}\left[  g,\Psi\right]  , \label{GR action}%
\end{equation}
where $S_{mat}$ represents the matter sector of the theory, $\Psi$ the matter
fields, and $\Lambda$ the cosmological constant. The gravitational sector, ie
$\frac{1}{16\pi G}\int\sqrt{-g}\left(  R-2\Lambda\right)  d^{4}x$ (the
Einstein-Hilbert action, here completed with a cosmological term), depends on
the spacetime metric only. Varying (\ref{GR action}) w.r.t. the metric yields
the GR Einstein's equation%
\begin{equation}
R_{ab}-\frac{1}{2}Rg_{ab}+\Lambda g_{ab}=8\pi GT_{ab}. \label{GR Einstein eq}%
\end{equation}
The variation w.r.t. $\Psi$ yieds the matter equations.

The lagrangian of the ST theory reads, in Jordan's representation [14][3]%
\begin{equation}
S_{ST}\left[  g,\Psi\right]  =\frac{1}{16\pi}\int\left(  \Phi R-\frac
{\omega\left(  \Phi\right)  }{\Phi}g^{ab}\partial_{a}\Phi\partial_{b}%
\Phi-2\Phi U\left(  \Phi\right)  \right)  \sqrt{-g}d^{4}x+S_{m}\left[
g,\Psi\right]  . \label{ST action}%
\end{equation}
In this representation, the gravitational sector (the r.h.s. discarding the
$S_{m}$\ term) involves both the spacetime metric and a scalar field $\Phi$.
On the other hand, the scalar does not enter the matter sector. This justifies
our representation choice: test particles then describe geodesics, even in the
vacuum case\footnote{One could split $S_{m}$\ in two parts: $S_{m}%
=S_{m(active)}+S_{m(test)}$, where $S_{m(test)}$ describes the physics of some
(the ones the behaviour of which we are interested in) test particles, and
$S_{m(active)}$ includes all the matter that effectively curves spacetime.
Setting $S_{m(active)}=0$\ then leads to the vacuum equations of the theory.
But the test particles' behaviour remain fully determined by $S_{m(test)}$.
Supposing that $S_{m}$ is globally scalar independent unambiguously results in
the fact that test particles describe geodesics, even if vacuum solutions are
considered.}. Each specific choice of the scalar functions $\omega\left(
\Phi\right)  $ and $U\left(  \Phi\right)  $ defines a specific ST theory.
Varying (\ref{ST action}) w.r.t. the metric yields the ST Einstein's equation
[3]%
\begin{equation}
\Phi\left(  R_{ab}-\frac{1}{2}Rg_{ab}+Ug_{ab}\right)  =8\pi T_{ab}%
+\frac{\omega}{\Phi}\left(  \partial_{a}\Phi\partial_{b}\Phi-\frac{1}{2}%
g_{ab}g^{cd}\partial_{c}\Phi\partial_{d}\Phi\right)  +\nabla_{a}\partial
_{b}\Phi-g_{ab}\square\Phi. \label{ST Einstein eq}%
\end{equation}
Varying (\ref{ST action}) w.r.t. the scalar field, and eliminating the Ricci
scalar $R$ thanks to the contracted version of (\ref{ST Einstein eq}), one
gets the scalar equation%
\begin{equation}
\square\Phi=\frac{8\pi T-\omega^{\prime}\partial_{c}\Phi\partial^{c}\Phi
-2\Phi\left(  U-\Phi U^{\prime}\right)  }{2\omega+3} \label{ST scalar eq}%
\end{equation}
where $\omega^{\prime}\equiv\frac{d\omega}{d\Phi}$ and $U^{\prime}\equiv
\frac{dU}{d\Phi}$. The special case $U^{\prime}=0$, but with $U\neq0$,
corresponds to ST theory with a cosmological constant. The special case
$\omega^{\prime}=0$ corresponds to the BD theory (with a potential $U$). The
original version of the BD theory was formulated in the potential free case
[15]. Let us point out that requiring $\omega\left(  \Phi\right)  >-3/2$
ensures the Ostrogradskian stability of the theory [16]. In fact, we will even
be a bit more restrictive in this paper, just considering theories with
$\omega>0$. This choice will simplify some aspects of the coming discussions,
and will prove to be adapted to the purpose of the paper.

If the scalar field $\Phi$ is replaced by a constant $\Phi_{0}$, the equation
(\ref{ST Einstein eq}) reduces to the GR's equation (\ref{GR Einstein eq})
with $G=1/\Phi_{0}$ and $\Lambda=U\left(  \Phi_{0}\right)  $. But of course,
to get an ST solution, the equation (\ref{ST scalar eq}) has to be satisfied
too. For finite values of $\omega$ and $\omega^{\prime}$, the existence of a
solution with a constant scalar then requires%
\begin{equation}
\Phi_{0}\left[  U\left(  \Phi_{0}\right)  -\Phi_{0}U^{\prime}\left(  \Phi
_{0}\right)  \right]  =4\pi T. \label{cond for GR sol is ST sol}%
\end{equation}
Thence, unless the relation (\ref{cond for GR sol is ST sol}) occurs, the ST
theory does not admit spacetimes with constant scalar\footnote{Let us remark
that the possibility for GR like solutions with $G=1/\Phi_{0}$ in the case
where (\ref{cond for GR sol is ST sol}) occurs does not contradict that the
effective local gravitational constant reads $G_{eff}=\left(  2\omega
_{0}+4\right)  \left(  2\omega_{0}+3\right)  ^{-1}/\Phi_{0}$ (and not simply
$G_{eff}=1/\Phi_{0}$) where $\omega_{0}=\omega\left(  \Phi_{0}\right)  $.
Indeed, $G_{eff}$ results from (1) the $\left(  00\right)  $ Einstein's BD
equation (with no potential) in the weak field, slow motion and
quasi-stationary approximations, (2) the elimination of $\square\Phi$ thanks
to the scalar BD equation, (3) the reduction of the perfect fluid stress
tensor to $T_{ab}=\rho\delta_{0a}\delta_{0b}$ under the previous hypotheses,
leading to the Poisson like equation $\triangle V=-4\pi\left(  2\omega
_{0}+4\right)  \left(  2\omega_{0}+3\right)  ^{-1}\Phi_{0}^{-1}\rho$, where
$V$\ is the Newtonian potential [20]. But in the same time, the equation
(\ref{cond for GR sol is ST sol}) reduces to $\rho=0$ (since there is no
potential). In this condition, $\triangle V=0$, and defining the effective
Newton's constant is meaningless.}.

\bigskip

\noindent\textbf{II-2 - Brans's Class I spherical solution}

The complete stationary solution of the BD equations with no potential ($U=0$)
is known in the spherical case [15][17]. It is made of four families, usually
refered to as Brans's Class I, II, III and IV solutions. We specify here to
the Class I family. It can be written\footnote{The correspondance between this
form and the form given in [17] reads $\left(  s,\epsilon\right)  =\left(
\frac{C}{l},\frac{l\left(  C+2\right)  }{\left\vert l\left(  C+2\right)
\right\vert }\right)  $ and $\left(  l,C\right)  =\frac{\left(  2,2s\right)
}{-s+\epsilon\sqrt{4-\left(  3+2\omega\right)  s^{2}}}$ (notations of ref
[17], with $\lambda$ replaced by $l$).}%
\begin{align}
\Phi &  =\left(  \frac{r-k}{r+k}\right)  ^{s}\label{BD Class I solution}\\
ds^{2}  &  =-\left(  \frac{r-k}{r+k}\right)  ^{-s+2\epsilon\lambda}%
dt^{2}+\left(  \frac{r+k}{r}\right)  ^{4}\left(  \frac{r-k}{r+k}\right)
^{2-s-2\epsilon\lambda}\left(  dr^{2}+r^{2}d\theta^{2}+r^{2}\sin^{2}\theta
d\phi^{2}\right) \nonumber
\end{align}
where $\lambda$ is defined by%
\begin{equation}
\lambda=\frac{1}{2}\sqrt{4-\left(  3+2\omega\right)  s^{2}}\in\left[
0,1\right]  . \label{BD lambda def}%
\end{equation}
In (\ref{BD Class I solution}), $\epsilon$ is a sign, and $s$ and $k$ are two
arbitrary constants, but with $s$ satisfying%
\begin{equation}
\left\vert s\right\vert \leq\frac{2}{\sqrt{3+2\omega}}
\label{BD Class I existence condition}%
\end{equation}
for (\ref{BD lambda def}) to get sense.

The solutions $\left(  s,k,\epsilon\right)  $ and $\left(  -s,-k,-\epsilon
\right)  $ being the same, let us choose $\epsilon=+$\ once for all. We will
also consider that $k\neq0$, $k=0$ being the Minkowski spacetime.

Because of its nice geometrical interpretation, it is worth defining the areal
radius, that reads, from (\ref{BD Class I solution}) (with $\epsilon=+ $)%
\begin{equation}
R\left(  r\right)  =r\left(  \frac{r+k}{r}\right)  ^{2}\left(  \frac{r-k}%
{r+k}\right)  ^{1-\frac{s}{2}-\lambda}. \label{BD areal radius def}%
\end{equation}

For large values of $r$, the $g_{00}$ component of (\ref{BD Class I solution})
expands as%
\begin{equation}
g_{00}=-1+\frac{2\left(  2\lambda-s\right)  k}{r}+O\left(  \frac{1}{r^{2}%
}\right)  . \label{BD time-time exp}%
\end{equation}
Thence, if it is positive, the quantity%
\begin{equation}
m=\left(  2\lambda-s\right)  k \label{BD mass}%
\end{equation}
is the newtonian mass of the gravitational field, as measured by a far observer.

The solution (\ref{BD Class I solution}) is the BD counterpart of the GR
Schwarzschild solution. However, while the physical meaning of the
Schwarzschild solution (as well as its Kruskal maximal extension) is well
understood in terms of BH spacetime, the interpretation of the Brans's
solution (\ref{BD Class I solution}) is more complex, generically describing
NakS or wormhole spacetimes depending on the $\left(  s,\epsilon\right)  $ (or
on Brans's $\left(  l,C\right)  $) parameters [18][19][20]. For any (finite)
$\omega$, the minimally scalarized case $s=0$ corresponds to the GR
Schwarzschild's metric in isotropic coordinates. (If $U=0$, Schwarzschild is
indeed a peculiar BD vacuum solution whatever the \textit{finite} value of
$\omega$, in accordance with (\ref{cond for GR sol is ST sol}).)

In the following, we specify to positive mass NakS like solutions. From
(\ref{BD mass}), the positive mass condition reads%
\begin{equation}
\left(  2\lambda-s\right)  k>0. \label{positive mass cond}%
\end{equation}
Coming from $r=+\infty$ regions, the first singularity one meets is located at
$r=\left\vert k\right\vert $. The condition (\ref{positive mass cond}) leads
to%
\begin{equation}
g_{00}\left(  r=\left\vert k\right\vert \right)  =0
\label{infinite redshift eq}%
\end{equation}
which means that this singularity is an infinite redshift location. Discarding
the Schwarzschild case, its areal radius is either zero or infinite. The
infinite case corresponds to a wormhole like structure (that corresponds to
the case discussed in [21], section II-A-3). Let us discard this case and
demand the areal radius to be zero. From (\ref{BD areal radius def}), this
requires%
\begin{equation}
\left(  2-s-2\lambda\right)  k>0. \label{NakS cond}%
\end{equation}
For $k>0$, and thanks to $\lambda\in\left[  0,1\right]  $,
(\ref{positive mass cond}) and (\ref{NakS cond}) are fulfilled iff%
\begin{equation}
s\in\left[  \frac{-2}{\sqrt{3+2\omega}},0\right[  \cup\left]  \frac
{2}{2+\omega},\frac{\sqrt{2}}{\sqrt{2+\omega}}\right[  .
\label{s interv for m>0}%
\end{equation}
The corresponding $\lambda$ intervals are respectively%
\begin{align}
\lambda\left(  s<0\right)   &  \in\left[  0,1\right[
\label{lambda interv for m>0}\\
\lambda\left(  s>0\right)   &  \in\left]  \frac{1}{\sqrt{2\left(
2+\omega\right)  }},\frac{1+\omega}{2+\omega}\right[  .\nonumber
\end{align}
As this will prove to be sufficient for the purpose of this paper, we will
just consider solutions with $k>0$. (\ref{s interv for m>0}) is then a
necessary and sufficient condition for the solution to represent a positive
mass NakS spacetime (given that we have restricted our study to BD with
$\omega>0$). Accordingly, the spacetime region with $r>k$ is considered.

\bigskip

\noindent\textbf{II-3 - Circular orbits in a Brans's NakS spacetime}

Massless ($\sigma=0$) circular orbital radii solve, from
(\ref{geod first integral}) and the second of (\ref{geod eq})%
\begin{equation}
\left(  \frac{g_{00}}{g_{22}}\right)  ^{\prime}=0.
\label{massless circ orb eq}%
\end{equation}
Replacing by the metric functions entering (\ref{BD Class I solution}) with
$\epsilon=+$, one gets the circular radius as a function of $s$. Since $r>k>0
$, it reads, using the mass (\ref{BD mass})%
\begin{equation}
r_{0}\left(  s\right)  =\frac{2\lambda+\sqrt{4\lambda^{2}-1}}{2\lambda-s}m.
\label{app A2}%
\end{equation}
If the areal radius (\ref{BD areal radius def}) is prefered to $r$, it reads%
\begin{equation}
R_{0}\left(  s\right)  =\frac{m}{2\lambda-s}\frac{\left(  2\lambda
+\sqrt{4\lambda^{2}-1}+1\right)  ^{2}}{2\lambda+\sqrt{4\lambda^{2}-1}}\left(
\frac{2\lambda+\sqrt{4\lambda^{2}-1}-1}{2\lambda+\sqrt{4\lambda^{2}-1}%
+1}\right)  ^{1-\frac{s}{2}-\lambda}. \label{BD massless areal radius}%
\end{equation}

Let us now consider massive circular orbits ($\sigma=1$). Eliminating $\left(
dt/dp\right)  ^{2}$ from (\ref{geod first integral}) and the second of
(\ref{geod eq}), and eliminating proper time derivatives thanks to the third
of (\ref{geod eq}), one gets the Binet equation%
\begin{equation}
2\frac{d^{2}r}{d\theta^{2}}+\left(  \ln\left\vert \frac{g_{00}}{r^{2}g_{22}%
}\right\vert \right)  ^{\prime}\left(  \frac{dr}{d\theta}\right)  ^{2}%
=r^{2}\left[  \left(  \ln g_{22}\right)  ^{\prime}-\left(  \frac{g_{22}}%
{C^{2}}+1\right)  \left(  \ln\left\vert g_{00}\right\vert \right)  ^{\prime
}\right]  \label{Binet}%
\end{equation}
where one has used (\ref{isotropic condition}). The radius $r_{circ}$\ of a
circular orbit with areal constant $C$ then solves%
\begin{equation}
\left[  \left(  \ln g_{22}\right)  ^{\prime}-\left(  \frac{g_{22}}{C^{2}%
}+1\right)  \left(  \ln\left\vert g_{00}\right\vert \right)  ^{\prime}\right]
_{\left(  r=r_{circ}\right)  }=0. \label{Binet circ orb eq}%
\end{equation}
Let us now consider an orbit close to the circular one%
\begin{equation}
r=r_{circ}+\rho\text{ \ \ with \ \ }\left\vert \rho\right\vert <<r_{circ}.
\label{radius pert def}%
\end{equation}
Linearizing (\ref{Binet}) w.r.t. $\rho$, one gets%
\begin{equation}
\frac{d^{2}\rho}{d\theta^{2}}=\frac{r_{circ}^{2}}{2}\left[  \left(  \ln
g_{22}\right)  ^{\prime}-\left(  \frac{g_{22}}{C^{2}}+1\right)  \left(
\ln\left\vert g_{00}\right\vert \right)  ^{\prime}\right]  _{\left(
r=r_{circ}\right)  }^{\prime}\rho+\frac{C^{\prime}-C}{C^{3}}r_{circ}%
^{2}\left[  g_{22}\left(  \ln\left\vert g_{00}\right\vert \right)  ^{\prime
}\right]  _{\left(  r=r_{circ}\right)  } \label{linearized Binet}%
\end{equation}
$C^{\prime}$ being the areal constant on the perturbed orbit. Thence, a radius
that solves%
\begin{equation}
\left[  \left(  \ln g_{22}\right)  ^{\prime}-\left(  \frac{g_{22}}{C^{2}%
}+1\right)  \left(  \ln\left\vert g_{00}\right\vert \right)  ^{\prime}\right]
^{\prime}=0 \label{stability limit 1}%
\end{equation}
separates stable circular orbits from unstable ones (thence generalizes the GR
ISCO). Eliminating $C^{2}$\ thanks to (\ref{Binet circ orb eq}), one gets%
\begin{equation}
\left(  \frac{\left(  \ln\left\vert \frac{g_{22}}{g_{00}}\right\vert \right)
^{\prime}}{g_{22}\left(  \ln\left\vert g_{00}\right\vert \right)  ^{\prime}%
}\right)  ^{\prime}=0. \label{stability limit 2}%
\end{equation}
Replacing by the metric functions entering (\ref{BD Class I solution}), one
gets a quartic equation on $r$, that achieves the form%
\begin{equation}
\left(  \frac{r}{k}+\frac{k}{r}-6\lambda-s\right)  ^{2}=4\left(  5\lambda
^{2}-1\right)  +\left(  s+4\lambda\right)  s. \label{stability limit 3}%
\end{equation}
It leads to the two ($\epsilon_{r}=\pm$) second order equations%
\begin{equation}
\left(  \frac{r}{k}-H_{\epsilon_{r}}\right)  ^{2}=H_{\epsilon_{r}}^{2}-1
\label{stability limit 4}%
\end{equation}
with%
\begin{equation}
H_{\epsilon_{r}}=3\lambda+\frac{s}{2}+\frac{\epsilon_{r}}{\sqrt{5}}%
\sqrt{\left(  5\lambda+\frac{s}{2}\right)  ^{2}-5+4\frac{1-\lambda^{2}%
}{3+2\omega}}. \label{Heps def}%
\end{equation}
A first condition of existence of solutions requires the r.h.s. of
(\ref{stability limit 3}), or equivalently the argument of the square root
entering (\ref{Heps def}), to be positive. A second condition reads
$H_{\epsilon_{r}}^{2}\geq1$, from (\ref{stability limit 4}). Finally, for a
given $H_{\epsilon_{r}}$ satisying $H_{\epsilon_{r}}^{2}\geq1$, only one of
the two solutions of (\ref{stability limit 4}) is $>k$ (the product of the two
roots being $=k^{2}$).

\bigskip

\noindent\textbf{II-4 - The large }$\omega$\textbf{\ case}

Motivated by the current experimental context, let us now examine more
specifically the large $\omega$\ case. From
(\ref{BD Class I existence condition}), $s\longrightarrow0$ when
$\omega\longrightarrow\infty$. On the other hand, from (\ref{BD lambda def}),
$\lambda$ can achieve any value in $\left[  0,1\right]  $, even in the limit
$\omega\longrightarrow\infty$. Let us remark that, for both signs of $s$, and
discarding Schwarzschild (NakS solutions being considered), the positive mass
NakS condition (\ref{lambda interv for m>0}) becomes $\left]  0,1\right[  $ in
the limit, ie the full set of possible $\lambda$ with non-zero mass. This
means that the Brans's Class I solution generically describes a positive mass
(remind $k>0$) NakS spacetime for large $\omega$, and that wormhole like
solutions no longer exist in the limit.

The relevant limit equations are: (1) the metric, coming from
(\ref{BD Class I solution}), that reads (with $\epsilon=+$)
\begin{equation}
ds^{2}=-\left(  \frac{r-k}{r+k}\right)  ^{2\lambda}dt^{2}+\left(  \frac
{r+k}{r}\right)  ^{4}\left(  \frac{r-k}{r+k}\right)  ^{2-2\lambda}\left(
dr^{2}+r^{2}d\theta^{2}+r^{2}\sin^{2}\theta d\phi^{2}\right)
\label{BDlim lambda}%
\end{equation}
(2) the massless circular radius, coming from (\ref{BD massless areal radius}%
), that reads
\begin{equation}
R_{0}\left(  \lambda\right)  =\frac{m}{2\lambda}\frac{\left(  2\lambda
+\sqrt{4\lambda^{2}-1}+1\right)  ^{2}}{2\lambda+\sqrt{4\lambda^{2}-1}}\left(
\frac{2\lambda+\sqrt{4\lambda^{2}-1}-1}{2\lambda+\sqrt{4\lambda^{2}-1}%
+1}\right)  ^{1-\lambda} \label{BDlim massless areal radius}%
\end{equation}
(3) the "ISCO" radius equation, coming from (\ref{stability limit 4}) and
(\ref{Heps def}), that reads%
\begin{equation}
\left(  \frac{r}{k}-3\lambda-\epsilon_{r}\sqrt{5\lambda^{2}-1}\right)
^{2}=\left(  3\lambda+\epsilon_{r}\sqrt{5\lambda^{2}-1}\right)  ^{2}-1.
\label{BDlim ISCO areal radius}%
\end{equation}
The metric (\ref{BDlim lambda}) is a GR solution, but filled by a massless
scalar, as expected in the $\omega\longrightarrow\infty$ limit of vacuum BD
context.\ Indeed, a massless scalar, originating from the vanishing gradient
of the BD scalar, has to be added to the original stress tensor $T_{ab}$ for
the full richness of the large $\omega$ BD (filled by $T_{ab}$) to be restored
[22][23]. Explicitly, the metric (\ref{BDlim lambda}) solves $R_{ab}%
=\partial_{a}\varphi\partial_{b}\varphi$, with%
\begin{equation}
\varphi=\sqrt{2\left(  1-\lambda^{2}\right)  }\ln\left\vert \frac{r-k}%
{r+k}\right\vert . \label{BDlim massless scalar}%
\end{equation}
The mass of the field, coming from (\ref{BD mass}), reads%
\begin{equation}
m=2\lambda k. \label{BDlim lambda mass}%
\end{equation}
From (\ref{BDlim massless areal radius}), massless circular orbits exist for
$\lambda<1/2$ only.

The scalar (\ref{BDlim massless scalar}) vanishes for $\lambda=1$, as expected
since (\ref{BDlim lambda}) reduces to Schwarzschild then. On the other hand,
the metric of the maximally scalarized case ($\lambda=0$, that yields $m=0$)
reads
\begin{equation}
ds^{2}=-dt^{2}+\left(  1-\frac{k^{2}}{r^{2}}\right)  ^{2}\left(  dr^{2}%
+r^{2}d\theta^{2}+r^{2}\sin^{2}\theta d\phi^{2}\right)  .
\label{BDlim max scal}%
\end{equation}
This is the metric (2.7) of [21], which is then just one of the full set of
all possible limits. Let us remark that the coordinate change%
\begin{equation}
u=r+\frac{k^{2}}{r} \label{from Schw to JNW coord}%
\end{equation}
puts (\ref{BDlim max scal}) in the following "close to Minkowski" form%
\begin{equation}
ds^{2}=-dt^{2}+du^{2}+\left(  u^{2}-4k^{2}\right)  \left(  d\theta^{2}%
+\sin^{2}\theta d\phi^{2}\right)  . \label{BDlim max scal 2}%
\end{equation}
In some sense, the Schwarzschild solution and (\ref{BDlim max scal}) are the
two extreme cases that can be encountered in the $\omega\longrightarrow\infty
$\ limit of Brans's solution.

It appears that (\ref{BDlim lambda})-(\ref{BDlim massless scalar}) is nothing
but the Janis-Newman-Winicour (JNW) solution [24], that precisely solves the
GR equation filled by a massless scalar, up to a radial coordinate
transform\footnote{The link between the two forms is got making first the
shift $u=R+\frac{r_{0}}{2}$ (JNW notations), and then the coordinate change
(\ref{from Schw to JNW coord}).}. Circular orbits in this metric have been
studied in [25], and the BD limit equations (\ref{BDlim massless areal radius}%
) and (\ref{BDlim ISCO areal radius}) are in agreement with their results. In
our notations and coordinates, massive circular orbits exist for all
$r>$\ $r_{0}\left(  \lambda\right)  $ if $1/2<\lambda\leq1$, and for all $r$
up to the singularity if $0<\lambda<1/2$. All the circular orbits are stable
if $\lambda<1/\sqrt{5}$. There is an ISCO orbit for $\lambda>1/2$. For
$1/\sqrt{5}<\lambda<1/2$, there are two branches of orbits that separate
stable from unstable orbits, that join together at the point $\left(
\lambda,\frac{R}{m}\right)  =\left(  \frac{1}{\sqrt{5}},2\left(  \frac
{3+\sqrt{5}}{2}\right)  ^{\frac{1}{\sqrt{5}}}\right)  \simeq\left(
0.447\,21,3.\,\allowbreak075\,8\right)  $.

All these results strengthen that, in agreement with [22][23], the GR filled
by a massless scalar (instead of Schwarzschild) equations constitutes the
relevant framework to treat the large $\omega$ vacuum BD problem, with no loss
of the Full richness of large $\omega$\ BD solutions.

\bigskip

\noindent\textbf{III - Observed frequencies measured by a far observer}

The fact that stable circular orbits exist in BD gravity for all areal radius
is a salient difference with GR, for which such orbits exist for $R>6m $ only
($R>3m$ if the stability requirement is removed). Determining the radius
dependence of the corresponding orbital frequency that measures a far observer
is of utmost relevance. Indeed, this quantity masters a lot of observable
physical phenomena originating in such orbital motions.

Since experimental data favour ST gravity with $\omega\gtrsim4.10^{4}$ [3], we
will take advantage of the conclusions got in the previous section to make the
calculations using the metric (\ref{BDlim lambda}). We will then check that
the main conclusions remain valid for large, but finite, values of $\omega$,
with the help of the formulae got in II-3. We then discuss the incidences of
the results on the gravitational radiation by systems that can reasonably be
modelized by the Brans's solution.

\bigskip

\noindent\textbf{III-1 - Massless circular orbits}

The (proper) period on the massless circular orbit, measured by an observer at
rest at a point of the orbit, reads $2\pi R_{0}\left(  \lambda\right)  $,
since the orbit is travelled at the unit speed. The frequency measured by a
far observer is affected by the Einstein effect, and reads%
\begin{align}
\nu_{\infty}\left(  \lambda\right)   &  =\frac{1}{2\pi R_{0}\left(
\lambda\right)  }\sqrt{\frac{-g_{00}\left(  r=r_{0}\right)  }{-g_{00}\left(
r=\infty\right)  }}\label{massless orb freq}\\
&  =\frac{\lambda}{\pi m}\frac{2\lambda+\sqrt{4\lambda^{2}-1}}{\left(
2\lambda+\sqrt{4\lambda^{2}-1}+1\right)  ^{2}}\left(  \frac{2\lambda
+\sqrt{4\lambda^{2}-1}-1}{2\lambda+\sqrt{4\lambda^{2}-1}+1}\right)
^{2\lambda-1}\nonumber
\end{align}
since $g_{00}\left(  r=\infty\right)  =-1$. For $\lambda=1$, one recovers the
GR value $\left(  6\pi\sqrt{3}m\right)  ^{-1}$, as expected. Let us define $f$
by normalizing (\ref{massless orb freq}) by this GR value%
\begin{align}
f\left(  \lambda\right)   &  \equiv\frac{\nu_{\infty}\left(  \lambda\right)
}{\nu_{\infty}\left(  \lambda=1\right)  }\label{massless orb norm freq}\\
&  =6\sqrt{3}\lambda\frac{2\lambda+\sqrt{4\lambda^{2}-1}}{\left(
2\lambda+\sqrt{4\lambda^{2}-1}+1\right)  ^{2}}\left(  \frac{2\lambda
+\sqrt{4\lambda^{2}-1}-1}{2\lambda+\sqrt{4\lambda^{2}-1}+1}\right)
^{2\lambda-1}.\nonumber
\end{align}
For $\lambda\longrightarrow1/2$, $f$ gets the value%
\begin{equation}
f\left(  \lambda\longrightarrow\frac{1}{2}\right)  \longrightarrow\frac
{3\sqrt{3}}{4}\simeq1.299.\label{massless orb freq for 1/2}%
\end{equation}
Interestingly, this value is not infinite, despite the fact that the massless
orbital circumference is $2\pi R\left(  \lambda=1/2\right)  =0$. The local
frequency is then infinite, but it is also infinitely redshifted since the
massless orbits meets the NakS, that is an infinite redshift location, for
$\lambda=1/2$. The two effects compete and finally compensate each other,
resulting in the finite far observed frequency given by
(\ref{massless orb freq}).

For $\lambda>1/2$, $f$ decreases until the GR value $f\left(  \lambda
=1\right)  =1$.

\bigskip

\noindent\textbf{III-2 - Massive circular orbits}

On any circular orbit, the time needed to complete a revolution, as measured
by an observer at rest at a point of the orbit, reads%
\begin{equation}
\tau=\sqrt{-g_{00}\left(  r\right)  }\frac{2\pi}{d\theta/dt}.
\label{proper period at orbit}%
\end{equation}
The frequency, as measured by a far observer, is then%
\begin{equation}
\nu_{\infty}=\frac{1}{\tau}\sqrt{\frac{-g_{00}\left(  r\right)  }%
{-g_{00}\left(  r=\infty\right)  }}=\frac{1}{2\pi}\frac{d\theta}{dt}
\label{far away freq}%
\end{equation}
Using the second equation of (\ref{geod eq}) and the metric functions entering
(\ref{BDlim lambda}), one gets%
\begin{equation}
\nu_{\infty}\left(  \lambda,r\right)  =\frac{1}{2\pi}\frac{\left(
2\lambda\right)  ^{\frac{3}{2}}}{m}\left(  \frac{2\lambda r}{m}\right)
^{\frac{3}{2}}\frac{\left(  \frac{2\lambda r}{m}-1\right)  ^{2\lambda
-1}\left(  \frac{2\lambda r}{m}+1\right)  ^{-2\lambda-1}}{\sqrt{\left(
\frac{2\lambda r}{m}\right)  ^{2}-2\lambda\frac{2\lambda r}{m}+1}}.
\label{massive orb freq (r)}%
\end{equation}
For $\lambda=1$, one recovers the GR value that reads, in terms of the areal
radius $R$, $\nu_{\infty}\left(  \lambda=1,r\right)  =\left[  2\pi\left(
R/m\right)  ^{\frac{3}{2}}m\right]  ^{-1}$.

In the GR case, the maximum frequency on stable orbits corresponds to the ISCO
($R=6m$) and reads $\left(  12\pi\sqrt{6}m\right)  ^{-1}$. For $1/\sqrt
{5}<\lambda<1$, the frequency of the $\lambda$ dependent ISCO is got inserting
the relevant solution of (\ref{BDlim ISCO areal radius}) in
(\ref{massive orb freq (r)}). It increases when $\lambda$ decreases, and
reaches the value $\left(  \frac{7-3\sqrt{5}}{2}\right)  ^{1/\sqrt{5}}\left(
4\pi\sqrt{2}m\right)  ^{-1}\simeq2.197\times\left(  12\pi\sqrt{6}m\right)
^{-1}$ for $\lambda=1/\sqrt{5}$. For $\lambda=1/2$, the frequency of the ISCO
is $\left(  12\pi\sqrt{3}m\right)  ^{-1}\simeq1.414\times\left(  12\pi\sqrt
{6}m\right)  ^{-1}$.

For $\lambda<1/\sqrt{5}$, (\ref{BDlim ISCO areal radius}) has no solution and
$\lambda$-ISCO(s) no longer exist. In this case, circular orbits exist until
the naked singularity (zero areal radius), and are all stable. For $1/\sqrt
{5}<\lambda<1/2$, there are unstable circular orbits, but stable circular
orbits still exist close to the NakS. For $\lambda<1/2$,
(\ref{massive orb freq (r)}) behaves, when approaching the singularity, as%
\begin{equation}
\nu_{\infty}\left(  \lambda,r\longrightarrow\frac{m}{2\lambda}\right)
\sim\frac{1}{2\pi m}\frac{\left(  2\lambda\right)  ^{\frac{3}{2}}}%
{2^{2\lambda+1}\sqrt{2\left(  1-\lambda\right)  }}\left(  \frac{2\lambda r}%
{m}-1\right)  ^{-\left(  1-2\lambda\right)  }.
\label{massive orb freq at singu}%
\end{equation}
This quantity is unbounded for all $\lambda<1/2$. As in the massless case, the
infinite local proper frequency competes with the infinite redshifting, but
unlike the massless case, the local proper frequency now dominates the
behaviour. As it should be, the massless case is nevertheless back using
(\ref{massive orb freq at singu}), since the massless circular orbit and the
singularity meet for $\lambda=1/2$. Indeed, for $\lambda=1/2$, the divergent
term is killed and (\ref{massive orb freq at singu}) yields $\nu_{\infty
}=\left(  8\pi m\right)  ^{-1}$, in accordance with
(\ref{massless orb freq for 1/2}).

\bigskip

\noindent\textbf{III-3 - Far frequency for large, but finite, }$\omega$

The fact that the far measured frequency (\ref{massive orb freq at singu}) on
stable circular orbits is unbounded for decreasing radius in the $\lambda<1/2$
case is an appealing difference with GR, where this quantity is bounded by the
ISCO value $\left(  12\pi\sqrt{6}m\right)  ^{-1}$. For that reason, it is
worth to check that this behaviour is not an artefact of the limit, but still
occurs in BD theory, for which $\omega$ is actually finite, even if very high.

The far measured frequency is still given by (\ref{far away freq}), but with
$d\theta/dt$ calculated from the metric functions entering
(\ref{BD Class I solution}) (with $\epsilon=+$). From the second of
(\ref{geod eq}), one gets%
\begin{align}
\nu_{\infty}\left(  \omega,s,r\right)   &  =\frac{1}{2\pi}\sqrt{\frac
{-g_{00}^{\prime}}{g_{22}^{\prime}}}\label{BD massive orb freq(r)}\\
&  =\frac{1}{2\pi}\left(  r-k\right)  ^{2\lambda-1}\left(  r+k\right)
^{-2\lambda-1}\sqrt{\frac{\left(  2\lambda-s\right)  kr^{3}}{r^{2}-\left(
s+2\lambda\right)  kr+k^{2}}}.\nonumber
\end{align}
For $r\longrightarrow k$, one gets the behaviour%
\begin{equation}
\nu_{\infty}\left(  \omega,s,r\longrightarrow k\right)  \longrightarrow
\frac{1}{2^{2\lambda+2}\pi k}\sqrt{\frac{2\lambda-s}{2-s-2\lambda}}\left(
\frac{r}{k}-1\right)  ^{2\lambda-1} \label{BD massive orb freq at singu}%
\end{equation}
(which leads back to (\ref{massive orb freq at singu}) for $\omega
\longrightarrow\infty$). The square root in front of
(\ref{BD massive orb freq at singu}) is defined thanks to
(\ref{positive mass cond}) \& (\ref{NakS cond}). Obviously,
(\ref{BD massive orb freq at singu}) diverges for $\lambda<1/2$, as
(\ref{massive orb freq at singu}) did. To be complete, let us remark that, for
all positive $\omega$, the quantity%
\[
8-\left(  7+5\omega\right)  s^{2}+s\sqrt{4-\left(  3+2\omega\right)  s^{2}}%
\]
is negative for values of $s^{2}$ sufficiently close to $4/\left(
3+2\omega\right)  $. This means that the square root entering (\ref{Heps def})
is negative then, in such a way that (\ref{stability limit 3}) has no
solution. The stability of the circular orbits is then ensured whatever their
radii for these values of $s$.

\bigskip

\noindent\textbf{III-4 - Physical impacts and gravitational radiation by EMRI
binaries in Brans-Dicke gravity}

The main objection usually raised against NakSs is that they generate
unpredictability in the parts of the Universe belonging to their causal
future. Thence the requirement to hide singularities behind horizons, to
preserve the part of the Universe we are belonging to from such an unwelcome
behaviour (Penrose's conjecture). However, the unpredictable behaviour
attached to a (naked or hidden) singularity is deeply related to the classical
character of the considered gravity theories. It is commonly believed that
singular states should be removed when the quantum nature of the spacetime
will be properly taken into account. In this sense, the presence of a (naked
or hidden) singularity is nothing but the hallmark of the fact that one enters
a region where the classical approximation of the "true" (presumably quantum)
gravity theory is no longer valid.

The possibility of NakSs being accepted along these lines, discussing the
potential astronomical applications of the solutions previously described
makes sense. The peculiar features with respect to GR got in the previous
sections result in potential differences in the expected behaviour of physical
observables that originate in strong gravitational fields. Some of them could
turn out as being of the utmost importance in the present astrophysical context.

For instance, the existence or not of an ISCO is expected to drastically
impact the accretion disk behaviour. This point has been discussed in details
in the JNW metric framework [25]. From the results here highlighted on the
large $\omega$ BD case, the conclusions got by these authors apply to ST
theories that are admissible in the current experimental context. Let us also
mention that similar phenomena have been observed in the framework of the
$q$-metric (also named $\gamma$-metric), which is another kind of spacetime
exhibiting NakS (while not related to ST gravity)[25][26].

The existence or not of a massless circular orbit obviously impacts
geometrical optics, at least for light rays penetrating strong field regions.
Indeed, in the GR framework, there is an infinite number of massless geodesics
circling the "photonic sphere", resulting in an infinite multiplication of
images (both of background and foreground objects) stuck on this sphere. This
feature is lost if there is no massless circular orbit. Specifying to a large
$\omega$ BD solution of mass $m$, the bending is weaker than in GR.\ This
means that the lensing object gravitational mass would be underestimated if
calculated on the ground of a given bending angle interpreted in the framework
of GR instead of BD. However, let us stress that the differences occur only
for orbits entering spacetime regions with $r/m$ of the order of some units.
Indeed, the $\gamma_{Edd}$ Eddington parameter of (\ref{BDlim lambda}) is
$=1$, meaning that geometrical optics does not differ from GR in the far
regions of the metric. Therefore, it seems unlikely that this effect could
significantly participate to the "missing mass" problem.

In the current astrophysical context, the most obvious impact probably
concerns the new GW astronomy. The unboundedless character of the orbital, and
then of the gravitational, frequency for some values of the parameter
$\lambda$ has no equivalent in the GR framework, where the far measured
gravitational frequency is bounded by its ISCO value $\left(  6\pi\sqrt
{6}m\right)  ^{-1}$ (twice the orbital frequency). Several works have already
been achieved on GW emission in the ST framework, but it is generally
presupposed that the central object is a (supermassive) BH [27], or the study
deals with Post-Newtonian approximation at some order [28]. Considering the
strong field region of a NakS like solution is clearly out of the scope of
these studies, and then requires a specific approach. Taking advantage of the
current experimental constraints on ST gravity, the solution
(\ref{BDlim lambda}) defines the natural framework to get the frequency and
amplitude of the GW emitted by a body of mass $\mu$ inspiraling a "non
rotating" BD like NakS of mass $m$, if $\mu<<m$ (extreme mass ratio
inspiraling --EMRI-- approximation)\footnote{Let us remind that, in ST
gravity, for the mass $\mu$\ body to describe a geodesics of the mass
$m$\ spacetime not only requires $\mu<<m$, but also the mass $\mu$\ body self
gravitational effects to be negligeable.}. The Brans's Class I metric serving
as the background solution, the orbital velocity, as measured by a local
observer, reads%
\begin{equation}
V\left(  \lambda,r\right)  =\sqrt{\frac{m}{r-m+\frac{m^{2}}{4\lambda^{2}r}}}.
\label{local velocity}%
\end{equation}
For a given value or the areal radius $R$, this velocity decreases when
$\lambda$ decreases\footnote{Indeed, $\left(  \frac{\partial\left(
V^{2}\right)  }{\partial\lambda}\right)  _{R}=\left(  \frac{\partial\left(
V^{2}\right)  }{\partial r}\right)  _{\lambda}\left(  \frac{\partial
r}{\partial\lambda}\right)  _{R}+\left(  \frac{\partial\left(  V^{2}\right)
}{\partial\lambda}\right)  _{r}>0$ since $\left(  \frac{\partial\left(
V^{2}\right)  }{\partial r}\right)  _{\lambda}<0 $ and $\left(  \frac
{\partial\left(  V^{2}\right)  }{\partial\lambda}\right)  _{r}>0$ from
(\ref{local velocity}), and $\left(  \frac{\partial r}{\partial\lambda
}\right)  _{R}<0$ from the $s=0$ version of (\ref{BD areal radius def}).}.
This suggests that, for a given areal radius $R$, the amplitude of the GW
should decrease for an increasing amplitude of the scalar
(\ref{BDlim massless scalar}). For $\lambda<1/2$, (\ref{local velocity}) is
bounded by the limit of its value when approaching the singularity, that reads
$\sqrt{\lambda/\left(  1-\lambda\right)  }$. The EMRI scheme should also
return the energy lost per revolution resulting from the GW radiation, giving
the adiabatic evolution of the orbit and of the frequency. If $\lambda>1/2$,
one could expect a behaviour qualitatively similar to GR, with and inspiraling
phase followed by a plunge, the differences just entering the numerical
details. If $\lambda<1/\sqrt{5}$, one could expect an inspiraling phase with
unbounded frequency (as long as only classical gravity, and no non
gravitational phenomena, enter the game). If $1/\sqrt{5}<\lambda<1/2$, a
plunge phase should occur after an inspiraling one, as in GR, but unlike GR, a
new inspiraling phase could occur at higher frequency and with higher
amplitude when the small body enters the inner stable circular orbits region.
In any cases, the process ends up with the "NakS hitting".

Since "NakS hitting" actually means that the small body enters a quantum
gravity region, it sounds reasonable to expect that this phase should be
encoded, in some way, in the corresponding part of the GW signal, offering an
opportunity to observe quantum gravity at work. However, as long as the
falling body remains far enough from the quantum region, classical
considerations should still be relevant. A rough estimate suggests that the
characteristic dimension of the quantum gravity region should be $\delta\sim
l_{Pl}\left(  m/m_{Pl}\right)  ^{1/3}\sim\left(  3.3\times10^{-28}%
\ \text{m}\right)  \left(  m/M_{Sun}\right)  ^{1/3}$, where $l_{Pl}$ and
$m_{Pl}$ are the Planck length and mass, and $M_{Sun}$ the mass of the Sun.
Comparing to the size $R$ of the critical orbits entering the figure 1, all of
the order of $m$, ie of $R\sim\left(  1.5\times10^{3}\ \text{m}\right)
\left(  m/M_{Sun}\right)  $, one gets%
\begin{equation}
\frac{\delta}{R}\sim\left(  2\times10^{-31}\right)  \left(  \frac{m}{M_{Sun}%
}\right)  ^{-2/3}. \label{expected QG region size}%
\end{equation}
Thence, for astrophysical expected masses, the quantum phenomena should enter
the game long after the falling body enters the internal stable orbits region
for $1/\sqrt{5}<\lambda<1/2$.

\bigskip

\noindent\textbf{IV - Conclusion}

The ability of solar system and classical tests to separate ST from GR
decreases for increasing $\omega$ [3]. In the case where the classical sector
of gravity in our Universe is of ST nature but with $\omega>$ (say) $10^{10}$,
this makes such tests unlikely to unveil the ST nature of gravity in a
foreseable future.

Much could change thanks to the new GW astronomy. Indeed, one has argued in
the present paper that if BD like NakS solutions do exist in our Universe (as
primordial objects, for instance), the GW signal generated by a small but
massive inspiraling body could drastically differ from GR models, in a way not
depending on how large $\omega$\ could be. This could result in very specific
ways to learn on the deep nature of gravity.

This should stimulate specific studies of GW emission by a massive body
orbiting the Brans like NakS solution. In the EMRI case, the Brans's solution
can serve as the background metric and scalar, in which the small but massive
body moves. Better: as argued in [22][23], a generic way to tackle a $T_{ab}$
filled BD (and, to some extent, ST, see [22]) problem in the large $\omega$
case is to consider GR instead, but filled by $T_{ab}$ plus an arbitrary
massless scalar field. The inclusion of this scalar, of finite amplitude, is a
necessary condition to restore the full $T_{ab}$ filled large $\omega$ BD
richness. This point has been more specifically illustrated in the vacuum case
in the present paper. Because of the current experimental constraints on
gravity theories, this approach is a natural way to tackle the binaries
evolution and radiation problem, especially in the EMRI case, since an exact
analytical BD solution can be used as the starting point then. This is an
alternative approach to existing works in ST gravity [27]\noindent\lbrack28],
which in any case are not adapted to the strong field NakS case.

\bigskip

\noindent\lbrack1] B. P. Abbott et al, Phys. Rev. Lett. \textbf{116}, 241102 (2016).

\noindent\lbrack2] B. P. Abbott et al, Phys. Rev. Lett. \textbf{116}, 241103 (2016).

\noindent\lbrack3] C.\ M.\ Will, \textit{The confrontation between general
relativity and experiments }in www.livingreviews.org/Irr-2014-4 (living
reviews in relativity).

\noindent\lbrack4] V.\ Faraoni, \textit{Cosmology in scalar-tensor gravity}
(Kluwer Academic Publishers, 2004).

\noindent\lbrack5] Y.\ Fujii, K. Maeda, \textit{The scalar-tensor theory of
gravitation} (Cambridge University Press, 2003).

\noindent\lbrack6] S.\ Capozziello, V. Faraoni, \textit{Beyond Einstein's
gravity}, Fundamental Theories of Physics, volume 170, Springer (2011).

\noindent\lbrack7] T.\ Damour, K.\ Nordtvedt, Phys.\ Rev.\ Lett. \textbf{70},
2217 (1993).

\noindent\lbrack8] T.\ Damour, K.\ Nordtvedt, Phys.\ Rev.\ D \textbf{48}, 3436 (1993).

\noindent\lbrack9] V. Cardoso, E. Franzin, P. Pani, Phys. Rev. Lett.
\textbf{116}, 171101 (2016); Phys. Rev. Lett. 117, 089902 (2016).

\noindent\lbrack10] V. Cardoso, S.\ Hopper, C.\ F.\ B.\ Macedo,
C.\ Palenzuela, P. Pani, Phys. Rev. D \textbf{94}, 084031 (2016).

\noindent\lbrack11] C.\ Chirenti, L.\ Rezzolla, Phys. Rev. D \textbf{94},
084016 (2016).

\noindent\lbrack12] N.\ Yunes, K.\ Yagi, F.\ Pretorius, Phys. Rev. D
\textbf{94}, 084002 (2016).

\noindent\lbrack13] C. W. Misner, K. S. Thorne, J. A. Wheeler,
\textit{Gravitation} (Freeman, San Francisco, 1973).

\noindent\lbrack14] C. M. Will, \textit{Theory and experiment in gravitational
physics} (Cambridge University Press, 1993).

\noindent\lbrack15] C. Brans and R. H. Dicke, Phys. Rev. \textbf{124}, 925 (1961).

\noindent\lbrack16] R.\ P.\ Woodard, Lecture Notes in Physics 720, 403 (2007).

\noindent\lbrack17] C. H. Brans, Phys. Rev. \textbf{125}, 2194 (1962).

\noindent\lbrack18] A. I. Janis, D. C. Robinson, J. Winicour, Phys. Rev.
\textbf{186}, 1729 (1969).

\noindent\lbrack19] A. G. Agnese, M. La Camera, Phys. Rev. D \textbf{51}, 2011 (1995).

\noindent\lbrack20] K. K. Nandi, A. Islam, J. Evans, Phys. Rev. D \textbf{55},
2497 (1997).

\noindent\lbrack21] V. Faraoni, F. Hammad, S. D. Belknap-Keet, Phys. Rev. D
\textbf{94}, 104019 (2016).

\noindent\lbrack22] B. Chauvineau, Class. Quant. Grav. \textbf{20}, 2617 (2003).

\noindent\lbrack23] B.\ Chauvineau, Gen. Rel. Grav. \textbf{39}, 297 (2007).

\noindent\lbrack24] A. I. Janis, E. T. Newman, J. Winicour, Phys. Rev. Lett.
\textbf{20}, 878 (1968).

\noindent\lbrack25] A.\ N.\ Chowdhury, M.\ Patil, D.\ Malafarina,
P.\ S.\ Joshi, Phys. Rev. D \textbf{85}, 104031 (2012).

\noindent\lbrack26] K.\ Boshkayev, E.\ Gasper\'{\i}n,
A.\ C.\ Guti\'{e}rrez-Pi\~{n}eres, H.\ Quevedo, S.\ Toktarbay, Phys. Rev. D
\textbf{93}, 024024 (2016).

\noindent\lbrack27] N.\ Yunes, P.\ Pani, V.\ Cardoso, Phys. Rev. D
\textbf{85}, 102003 (2012).

\noindent\lbrack28] N.\ Sennett, S.\ Marsat, A.\ Buonanno, Phys. Rev. D
\textbf{94}, 084003 (2016).
\end{document}